\documentclass[prb,reprint,superscriptaddress,showkeys,floatfix,nobalancelastpage]{revtex4-2}

\usepackage{graphicx}
\usepackage{color}
\usepackage{amsmath,amssymb}
\usepackage{braket}
\usepackage{bm}
\usepackage{comment}
\usepackage{upgreek}
\usepackage{hyperref}
\hypersetup
{
	colorlinks=true,
	allcolors=blue,
}

\newcommand{\fig}[1] {Fig.~\ref{#1}}
\newcommand{\figs}[1] {Figs.~\ref{#1}}
\newcommand{\figu}[1] {Figure~\ref{#1}}
\newcommand{\figus}[1] {Figures~\ref{#1}}
\newcommand{\eq}[1] {Eq.~\eqref{#1}}
\newcommand{\eqs}[1] {Eqs.~\eqref{#1}}
\newcommand{\equ}[1] {Equation~\eqref{#1}}
\newcommand{\x}[1] {\mathrm{#1}}
\newcommand{\tr}[1] { \x{Tr} \left\{ #1 \right\} }
\newcommand{\lr} {\lambda_\x{R}}
\newcommand{\li} {\lambda_\x{I}}
\newcommand{\lvz} {\lambda_\x{VZ}}
\newcommand{\neh} {N_\x{eh}}
\newcommand{\veh} {V_\x{eh}}
\newcommand{\leh} {l_\x{eh}}

\begin{document}

\title{Spin lifetime anisotropy in graphene induced by the SiO$_2$ interface}

\author{Aron W. Cummings}
\email{aron.cummings@icn2.cat}
\affiliation{Catalan Institute of Nanoscience and Nanotechnology (ICN2), CSIC and BIST, Campus UAB, Bellaterra, 08193 Barcelona, Spain}

\author{Chunhao Guo}
\thanks{A.W.C. and C.G. contributed equally.}

\author{Andrew Grieder}
\affiliation{Department of Materials Science and Engineering, University of Wisconsin-Madison, Madison, Wisconsin, 53706, United States}

\author{Shihao Tu}
\affiliation{Department of Materials Science and Engineering, University of Wisconsin-Madison, Madison, Wisconsin,  53706, United States}

\author{Mayank Gupta}
\affiliation{Department of Materials Science and Engineering, University of Wisconsin-Madison, Madison, Wisconsin, 53706, United States}

\author{Junqing Xu}
\altaffiliation[Current address: ]{Department of Physics, Hefei University of Technology, 420 Feicui Road, University City, Hefei Economic and Technological Development Zone, Anhui Province, China}
\affiliation{Department of Chemistry and Biochemistry, University of California, Santa Cruz, CA, 95060}

\author{Juan Marmolejo-Tejada}
\affiliation{Department of Electronics Engineering, Montana State University, 121 Montana Hall, Bozeman, MT 59717, United States}

\author{Yuan Ping}
\email{yping3@wisc.edu}
\affiliation{Department of Materials Science and Engineering, University of Wisconsin-Madison, Madison, Wisconsin, 53706, United States}
\affiliation{Department of Physics, University of Wisconsin-Madison, Madison, Wisconsin 53706, United States}
\affiliation{Department of Chemistry, University of Wisconsin-Madison, Madison, Wisconsin 53706, United States}

\begin{abstract}
Understanding how common dielectric substrates influence the spin transport properties of graphene is essential for advancing graphene-based spintronic technologies.
Here we use a comprehensive set of numerical simulations to reveal how a SiO$_2$ substrate modifies the spin texture and governs spin relaxation in graphene.
Using first-principles density matrix dynamics simulations, as well as tight-binding (TB) transport simulations, we quantify the effects of electron–phonon scattering, impurity scattering, and electrostatic disorder on the spin relaxation process.
We find that a 2D SiO$_2$ substrate induces a predominantly Rashba-type helical spin texture in graphene, leading to a spin lifetime anisotropy of 1/2.
Meanwhile, bulk SiO$_2$ breaks in-plane symmetry in graphene, leading to anisotropic in-plane and out-of-plane components in the spin texture, which we capture with a newly-developed TB model of graphene.
Transport simulations under realistic disorder conditions reveal a spin lifetime anisotropy between 0.5 and 1, similar to what is seen in measurements of graphene spin valves on a SiO$_2$ substrate.
Our results reveal a more complex picture of spin relaxation at the ubiquitous graphene/SiO$_2$ interface, beyond the standard Rashba model, providing critical insight for interpreting experiments and guiding substrate engineering for graphene spintronics.
\end{abstract}

\maketitle

\section{Introduction}

Graphene has proven to be a promising material for spintronic applications~\cite{Han2014_spinreview, Avsar2020_spinreview, Roche2024_spinreview}, owing to its high carrier mobility, weak intrinsic spin-orbit coupling (SOC), and negligible hyperfine interaction.
Such properties make graphene an ideal platform for the transport of spin information over long distances, with spin diffusion lengths reaching tens of $\upmu$m~\cite{Ingla2015, Drogeler2016, Gebeyehu2019, Panda2020, Bisswanger2022}.
However, despite being a good spin transporter, graphene on its own is a poor candidate for active spintronic devices, since the generation and manipulation of spin current requires large SOC or magnetism.
This limitation can be overcome by layering graphene with high-SOC or magnetic materials such as transition metal dichalcogenides (TMDCs), topological insulators, or magnetic insulators.
In such heterostructures, large SOC or magnetic exchange can be imprinted onto the graphene layer while maintaining its favorable charge transport properties~\cite{Gmitra2015grTMDC, Gmitra2016grTMDC, Song2018grTI, Zollner2018grFm}, leading to unique spin transport features~\cite{Cummings2017grTMDC, Ghiasi2017grTMDC, Benitez2018grTMDC}, and enabling new functionality for active graphene spintronic devices~\cite{Benitez2020grTMDC, Ghiasi2021grFM}.

As illustrated by the above examples, spin transport in graphene is highly susceptible to its surrounding environment, which includes the substrate on which it sits.
This raises the natural question, do common dielectric substrates such as SiO$_2$ also have a significant impact on spin transport in graphene?
This question becomes more important as the quality of graphene-based spintronic devices improves and extrinsic sources of defects and disorder are removed~\cite{Ingla2015, Drogeler2016, Gebeyehu2019, Panda2020, Bisswanger2022}.

For years, the understanding has been that SiO$_2$ breaks out-of-plane symmetry in graphene, which induces Rashba-like SOC and helical in-plane spin texture~\cite{Ertler2009, Huertas2009}.
This results in a spin lifetime anisotropy of $\zeta = 1/2$, where $\zeta \equiv \tau_{\x{s}z} / \tau_{\x{s}x}$ is the ratio of the lifetime of spins pointing out of the graphene plane to that of spins pointing in the plane~\cite{Dyakonov1972, Fabian2007}.
Initial experiments in nonlocal spin valves revealed nearly isotropic spin relaxation, $\zeta \approx 1$~\cite{Raes2016, Raes2017}, which could be explained by a dominant contribution from paramagnetic impurities~\cite{Kochan2014}, and suggests that the SiO$_2$ substrate plays no role in spin relaxation.
However, more recent measurements revealed $\zeta = 0.76-0.96$ depending on the measurement approach~\cite{Ringer2018}, suggesting that spin relaxation is driven by a mixture of spin-orbit effects, paramagnetic impurities, and contributions from ferromagnetic contacts~\cite{Zhu2018}.
These latest measurements hint at a contribution to spin relaxation arising from the SiO$_2$ substrate, but further work is needed to understand this in more detail.

We use a combination of first-principles density-matrix dynamics (FPDMD)~\cite{Xu2020, Xu2021, Xu2024-lw, Xu2024-ze, simoni2025} and tight-binding (TB) simulations to reexamine the impact of a SiO$_2$ substrate on spin transport in graphene.
We consider both 2D and bulk SiO$_2$ as substrates, and find distinct behavior for each.
For a 2D SiO$_2$ substrate, a standard Rashba-like spin texture is induced in graphene, and spin transport simulations reveal $\zeta = 1/2$.
Meanwhile, bulk SiO$_2$ breaks both out-of-plane and in-plane mirror symmetry in graphene, resulting in a spin texture that includes a nonuniform out-of-plane component.
Our spin transport calculations, under realistic disorder conditions, then reveal $0.5 < \zeta \lesssim 1$, similar to what has been measured experimentally~\cite{Raes2016, Raes2017, Ringer2018}.
Overall, our results show that the detailed nature of the graphene/SiO$_2$ interface can have a measurable impact on spin transport, and can yield behavior beyond the standard in-plane helical spin texture arising from Rashba SOC.
In particular, we find that the nearly isotropic spin relaxation measured in graphene spin valves can be driven by the SiO$_2$ substrate itself, in addition to any contributions from impurities or ferromagnetic contacts. Such behavior is expected to play a more significant, and eventually a limiting, role as graphene device quality continues to improve.

\section{Band structure and spin texture}

\subsection{First-principles calculations}

\begin{figure}[h]
	\centering
	\includegraphics[width=\columnwidth]{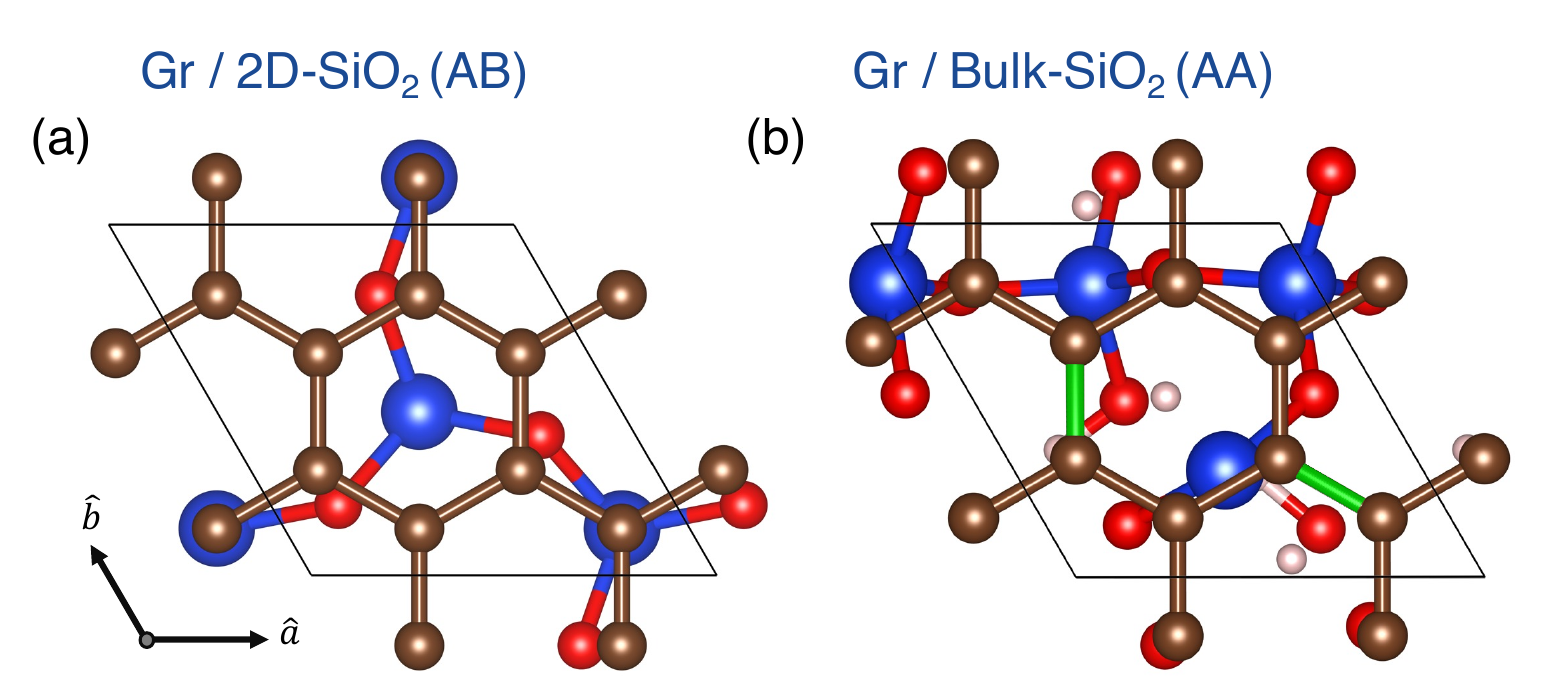}
	\caption{(a) Crystal structure of the graphene and 2D SiO$_2$ interface with AB stacking. (b) Crystal structure of graphene and bulk SiO$_2$ interface with AA stacking and Si termination. The green bonds in graphene indicate those modified by proximity to the SiO$_2$, see Eq.\ \eqref{eq_ham1} and the associated discussion.}
	\label{fig:structure}
\end{figure}

We begin by examining the electronic band structure and spin texture of graphene on SiO$_2$, which are essential to understand the spin relaxation mechanism and the spin lifetime anisotropy.
We consider 2D and bulk SiO$_2$ layers interfaced with graphene, as shown in \figs{fig:structure}(a) and (b), respectively.
Each structure consists of a $2 \times 2$ unit cell of graphene on a $1 \times 1$ unit cell of SiO$_2$.
We fixed the lattice constant of graphene but enforced lattice matching from SiO$_2$, to minimize the effect of strain on the electronic structure of graphene.
As seen in panel (a), the Si-O-Si bond of the 2D SiO$_2$ is relaxed to enable a commensurate alignment with the graphene layer.
This type of twisted Kagome lattice has previously been observed in 2D silica~\cite{bjorkman2016vibrational, doudin2022twisting}.
Here we have considered what we call an ``AB'' surface alignment where sublattice symmetry is broken in the graphene layer, as this is the lowest energy system.
In panel (b), the bulk SiO$_2$ is cut to a finite slab, and the surface is Si-terminated and passivated by OH adsorbates (more details on interface and slab structure can be found in the SM Sec.II).
We have considered several combinations of interface alignments and Si or O surface termination, and found this to be the lowest energy interface.

\begin{figure}[h]
	\includegraphics[width=\columnwidth]{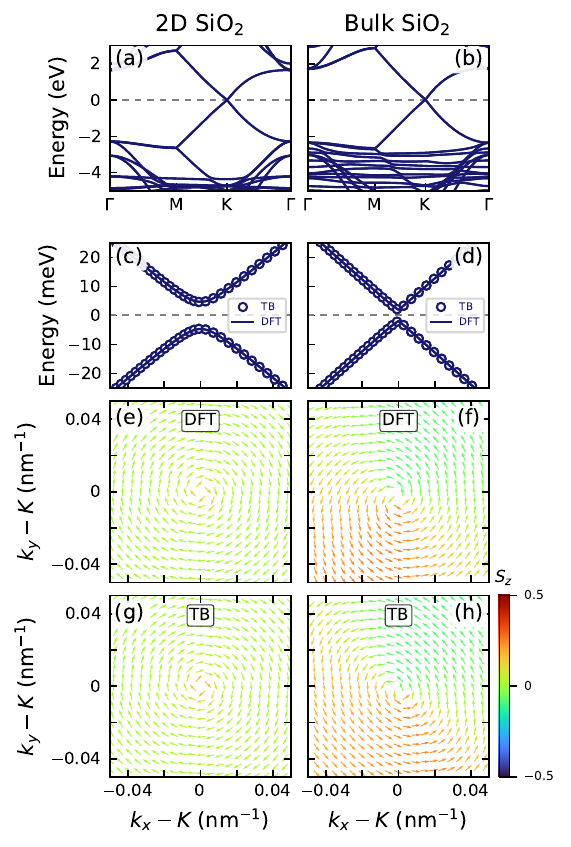}
	\caption{Band structure and spin texture of graphene on SiO$_2$. The left (right) column shows the case of graphene on 2D SiO$_2$ (bulk SiO$_2$). Panels (a) and (b) show the DFT band structure over the Brillouin zone, while (c) and (d) show a zoom of the bands around the graphene Dirac point, including the fit to the TB model. Panels (e) and (f) show the DFT spin texture of the the highest valence band of graphene, while (g) and (h) show the spin texture calculated from the TB model.}
	\label{fig:bs_st}
\end{figure}

The band structure and spin texture of the two systems are shown in \fig{fig:bs_st}.
These were calculated with the open-source \textit{ab initio} plane-wave code JDFTx~\cite{sundararaman2017jdftx}, using the Perdew-Burke-Ernzerhof (PBE) exchange-correlation functional along with fully relativistic optimized norm-conserving pseudopotentials~\cite{HamannONCV}, and a $k$-mesh of $24 \times 24 \times 1$ for the Brillouin zone integration.
The lattice constants and internal geometries were fully relaxed using the DFT+D3 method~\cite{GrimmeJCP-2010} for van der Waals dispersion corrections.

\figus{fig:bs_st}(a) and (b) show the DFT band structure of each system.
In both, the Dirac cone of graphene is preserved at the K point.
\figus{fig:bs_st}(c) and (d) highlight the band dispersion in the vicinity of the Dirac point, showing a small gap opening in both systems.
For graphene on bulk SiO$_2$, the apparent sharpness or discontinuity at the band edges indicates that the conduction-band minimum and valence-band maximum are laterally displaced in $k$-space from the K point.
This is further analyzed using the TB model discussed below.

\figus{fig:bs_st}(e) and (f) show the spin textures of the highest valence band of these systems, centered around the K point.
Starting with the graphene/2D-SiO$_2$ interface in \fig{fig:bs_st}(e), we see a purely helical in-plane spin texture with negligible out-of-plane spin polarization, consistent with broken out-of-plane symmetry and Rashba SOC induced in the graphene layer by the substrate~\cite{KaneMele2005grapheneSOC, Gmitra2009grapheneSOC}.
The small out-of-plane component near the K point is the result of the combination of Rashba SOC and sublattice breaking~\cite{Zollner2021grapheneSOC}.

Meanwhile, the spin texture of graphene on bulk SiO$_2$ in \fig{fig:bs_st}(f), while similar, exhibits key differences from a purely Rashba-like spin texture.
The first is that the in-plane spin texture is not purely helical, but rather shows a preferred orientation along the $k_y$ axis.
This is reminiscent of the persistent spin helix seen in 2DEGs in semiconductor quantum wells in the presence of both Rashba and Dresselhaus SOC~\cite{Winkler2004psh}.
Such a spin texture is expected to give rise to anisotropic in-plane spin lifetime~\cite{Averkiev1999aniso, Kainz2003aniso, Averkiev2006aniso}, as we will see below in our spin transport simulations, and has recently been measured in graphene on PdSe$_2$~\cite{Sierra2025graphenePdSe2}.
The second feature is the presence of a finite out-of-plane spin component.
Sublattice symmetry breaking in graphene can give rise to valley-Zeeman (VZ) SOC, which polarizes the spins out of the graphene plane with opposite direction in the K and K' valleys~\cite{Gmitra2016grTMDC, Cummings2017grTMDC}.
However, VZ SOC leads to a \textit{uniform} out-of-plane spin component in each valley, while in \fig{fig:bs_st}(f) we see a modulation and reversal of the spin with respect to the direction of $k$ within a single valley.

\subsection{Tight-binding model}

We now introduce a tight-binding model for graphene that captures the features of the band structure and spin texture seen \figs{fig:bs_st}(c)-(f).
Considering the same $2 \times 2$ unit cell as for the DFT calculations, we start with the standard Hamiltonian for graphene on a substrate~\cite{Gmitra2016grTMDC},
\begin{gather}
	\hat{H}_0 = t \sum_{\langle i,j \rangle} \hat{c}_{i}^\dagger \hat{c}_{j} + \frac{\Delta}{2} \sum_{i} \xi_i \hat{c}_{i}^\dagger \hat{c}_{i} \nonumber \\
	+ \frac{2\x{i}}{3} \lr \sum_{\langle i,j \rangle} \left(\hat{\bm{s}} \times \bm{d}_{ij} \right)_z  \hat{c}_{i}^\dagger \hat{c}_{j} \label{eq_ham0} \\
	+ \frac{\x{i}}{3\sqrt{3}} \sum_{\langle\langle i,j \rangle\rangle} \left( \li + \xi_i\lvz \right) \nu_{ij} \hat{s}_z \hat{c}_{i}^\dagger \hat{c}_{j}, \nonumber
\end{gather}
where $\hat{c}_i^\dagger$ $(\hat{c}_i)$ is the creation (annihilation) operator of an electron at carbon site $i$, while the single/double brackets indicate sums over first- and second-nearest neighbors, respectively.
The first term of the Hamiltonian is the nearest-neighbor hopping term with hopping energy $t$.
The second term, with $\xi_i = \pm 1$ for $i \in \{A,B\}$, breaks A/B sublattice symmetry and opens a band gap of $\Delta$.
The third term, where $\hat{\bm{s}}$ is the vector of Pauli spin operators and $\bm{d}_{ij}$ is the unit vector from site $j$ to $i$, is the Rashba SOC induced by the substrate.
This term induces the helical spin texture seen for example in \fig{fig:bs_st}(e).
The last term, where $\nu_{ij} = \pm 1$ for a clockwise (counterclockwise) hopping path between second neighbors $i$ and $j$, describes the standard Kane-Mele SOC that opens a topological gap of $2\li$~\cite{KaneMele2005grapheneSOC}, and the valley-Zeeman SOC that polarizes the spin out of the graphene plane with opposite sign in opposite valleys~\cite{Gmitra2016grTMDC, Cummings2017grTMDC}.

The Hamiltonian of \eq{eq_ham0} accounts for the primary features of graphene on a substrate, but it does not capture the effects observed in \fig{fig:bs_st}: the lateral $k$-shift of the band extrema and the anisotropic in-plane and out-of-plane spin textures.
To capture these effects, we take note of those atoms in the SiO$_2$ substrate that are nearest to the graphene layer.
These atoms are likely to break sublattice symmetry, as well as modify all the nearest-neighbor terms to which they are closest (highlighted as the green bonds in \fig{fig:structure}(b)).
We thus propose adding anisotropic terms to the Hamiltonian,
\begin{gather}
	\hat{H}_\x{aniso} = \sum_{\langle k,l \rangle} \delta t_{kl} \hat{c}_{k}^\dagger \hat{c}_{l} \nonumber \\
	+ \frac{2\x{i}}{3} \sum_{\langle k,l \rangle} \delta\lambda_{\x{R},kl} \left( \hat{\bm{s}} \times \bm{d}_{kl} \right)_z \hat{c}_{k}^\dagger \hat{c}_{l} \label{eq_ham1} \\
	+ 4\x{i} \sum_{\langle k,l \rangle} \lambda_{\x{R},kl}^\x{in} \hat{s}_z \hat{c}_{k}^\dagger \hat{c}_{l}. \nonumber
\end{gather}
Here the sums run over those nearest-neighbor bonds that are altered by the SiO$_2$ surface, as highlighted in \fig{fig:structure}(b).
The first term in \eq{eq_ham1} is the modification of the nearest-neighbor hopping, and the second term is the modification of the Rashba SOC along the impacted bonds.
The third term arises from broken in-plane mirror symmetry around the impacted bonds, which is equivalent to adding a local in-plane electric field.
This yields terms that couple to the $z$-component of the spin. As shown in the SM~\cite{SM}, the first term in \eq{eq_ham1} results in a lateral $k$-shift of the bands, the second term induces anisotropy in the in-plane helical spin texture, and the last term is responsible for the nonuniform out-of-plane spin texture seen in \fig{fig:bs_st}(f).

Next we fit the TB model of \eqs{eq_ham0} and \eqref{eq_ham1} to the DFT simulations over a chosen $k$-path.
A zoom of the fits to the band structure is shown in \figs{fig:bs_st}(c) and (d), and the full fits to the band structure and spin texture can be found in the SM~\cite{SM}.
Table \ref{tab_params} lists the fitted parameters.
From these parameters one can already see the difference between the two interfaces.
For graphene/2D-SiO$_2$, the standard Rashba SOC is dominant, while the anisotropic terms are all relatively small.
Meanwhile, in the graphene/bulk-SiO$_2$ interface, the anisotropic terms play a much larger role.
In \figs{fig:bs_st}(g) and (h) we plot the spin texture of the upper valence band from the TB model, as a comparison to the DFT results in the panels just above.
As seen here, this model captures the primary features found in the DFT simulations.

\begin{table}[t]
	\centering
	\begin{tabular*}{\columnwidth}{|@{\extracolsep{\fill} } c c c|}
		\hline
		Parameter & Gr/2D-SiO$_2$ & Gr/bulk-SiO$_2$ \\
		\hline\hline
		$t$ (eV)           & $2.47$  & $2.57$  \\
		$\Delta$ (meV)     & $4.69$  & $1$     \\
		$\lr$ ($\upmu$eV)  & $-9.41$ & $7.3$   \\
		$\li$ ($\upmu$eV)  & $0.57$  & $0.4$   \\
		$\lvz$ ($\upmu$eV) & $-0.03$ & $-0.69$ \\
		\hline
		$\delta t_{24}$ (meV)                   & $0.51$   & $-5.83$ \\
		$\delta t_{37}$ (meV)                   & $0.2$    & $1.43$ \\
		$\delta\lambda_{\x{R},24}$ ($\upmu$eV)  & $1.31$   & $-12.55$ \\
		$\delta\lambda_{\x{R},37}$ ($\upmu$eV)  & $-1.07$  & $26.4$ \\
		$\lambda_{\x{R},24}^\x{in}$ ($\upmu$eV) & $-0.063$ & $-0.75$ \\
		$\lambda_{\x{R},37}^\x{in}$ ($\upmu$eV) & $-0.045$ & $-1.85$ \\
		\hline
	\end{tabular*}
	\caption{Tight-binding parameters for the graphene Hamiltonian given in \eqs{eq_ham0} and \eqref{eq_ham1}.}
	\label{tab_params}
\end{table}

\section{Spin transport}

\subsection{Methodology}
\label{sec:methodology}

We now examine the nature of spin relaxation in these graphene/SiO$_2$ systems in the presence of electron-phonon coupling, defects, and disorder.
To accurately treat electron-phonon and electron-impurity scatterings, we use a first-principles technique based on Lindbladian dynamics of the one-particle electron density matrix~\cite{Xu2020abinitiodynamics, Xu2021abinitiodynamics, Xu2024-lw, Xu2024-ze}.
Specifically, we solve the quantum master equation of the density matrix $\hat{\rho}(t)$,
\begin{gather}
	\frac{\x{d}\hat{\rho}_{12}(t)}{\x{d}t} = -\frac{\x{i}}{\hbar} \left[ \hat{H}_\x{e} , \hat{\rho}(t) \right]_{12} \nonumber \\
	+ \left( \frac{1}{2} \sum_{345} \begin{bmatrix} (\hat{I}-\hat{\rho})_{13} \hat{P}_{32,45} \hat{\rho}_{45} \\ -(\hat{I}-\hat{\rho})_{45} \hat{P}_{45,13} \hat{\rho}_{32} \end{bmatrix} + \x{h.c.} \right),
	\label{eq:rho}
\end{gather}
where the numerical subscripts (e.g., ``1'') are combined $k$-point and band indices.
The first term in \eq{eq:rho} accounts for coherent dynamics of $\hat{\rho}$, with $\hat{H}_\x{e}$ the single-particle electronic Hamiltonian and external magnetic field perturbation, which can lead to spin-orbit and Larmor precession.
The second term captures scattering effects, where $\hat{I}$ is the identity matrix and $\hat{P}$ = $P^\x{e-ph}$ + $P^\x{e-imp}$ is a generalized scattering matrix that includes electron-phonon as well as electron-impurity scattering. The density matrices and Lindbladian scattering matrices are all computed from first-principles with Kohn-Sham states as basis. Spin-orbit couplings are self-consistently included through fully-relativisitic poseudopotentials~\cite{Xu2021}. 

Electron-phonon scattering is described via the scattering matrix
\begin{gather}
	\hat{P}_{1234}^{\x{e-ph}} = \sum_{\bm{q},\lambda,\pm} \hat{A}_{13}^{\bm{q} \lambda \pm} \left( \hat{A}_{24}^{\bm{q} \lambda \pm} \right)^*, \\
	\hat{A}_{13}^{\bm{q} \lambda \pm} = \sqrt{\frac{2\pi}{\hbar}} \hat{g}_{13}^{\bm{q} \lambda \pm} \sqrt{\delta_\sigma^\x{G}(\epsilon_1 - \epsilon_3 \pm \omega_{\bm{q}\lambda})} \sqrt{n_{\bm{q}\lambda}^{\pm}},
	\label{eq:sm_eph}
\end{gather}
where $\bm{q}$ and $\lambda$ are the phonon wave vector and mode, $g_{13}^{\bm{q} \lambda \pm}$ is the electron-phonon matrix element arising from phonon absorption ($+$) or emission ($-$) $n_{\bm{q}\lambda}^{\pm} = n_{\bm{q}\lambda} + 0.5 \pm 0.5$ with the phonon Bose occupation factor $n_{\bm{q}\lambda}$, and $\delta_\sigma^\x{G}$ is an energy-conserving delta function broadened to a Gaussian with a width of $\sigma$~\cite{Xu2020abinitiodynamics}.
Electron-phonon matrix elements are computed with finite differences~\cite{Xu2020}, using Wannier interpolations of high $k$ and $q$ meshes~\cite{Giustino2007-prb}, enabling efficient Brillouin zone convergence.

For electron-impurity scattering,
\begin{gather}
	\hat{P}_{1234}^\x{e-imp} = \hat{A}_{13}^\x{imp} \left( \hat{A}_{24}^\x{imp} \right)^*, \\ 
	\hat{A}_{13}^\x{imp} = \sqrt{\frac{2\pi}{\hbar}} \hat{g}_{13}^\x{imp} \sqrt{\delta_\sigma^\x{G}(\epsilon_1 - \epsilon_3)} \sqrt{n_\x{imp} V_\x{cell}},
	\label{eq:sm_imp}
\end{gather}
where $n_\x{imp}$ is the impurity density and $V_\x{cell}$ is the unit cell volume.
The electron-impurity matrix element is given by $\hat{g}_{13}^\x{imp} = \braket{1 | V^\x{imp} - V^0 | 3}$, where $V^\x{imp}$ ($V^0$) is the potential of the system with (without) the impurity computed using the supercell method~\cite{Xu2023, Xu2021abinitiodynamics}.

To understand the spin relaxation mechanism, it is also necessary to calculate the momentum relaxation times.
%
%
The momentum relaxation times of a given state $\ket{1}$ arising from electron-phonon and electron-impurity interactions are~\cite{Xu2024}
\begin{gather}
	\frac{1}{\tau_{p,1}^\x{e-ph}} = \frac{2\pi}{\hbar} \sum_2 \sum_{\bm{q} \lambda \pm} \left| \hat{g}_{12}^{\bm{q} \lambda \pm} \right|^2 \delta^G_\sigma(\epsilon_1 - \epsilon_2 \pm \omega_{\bm{q}\lambda}) n_{\bm{q}\lambda}^\pm \frac{\mathbf{v}_1 \cdot \mathbf{v}_2}{v_1v_2}, \label{eq:tp_ph} \\
	\frac{1}{\tau_{p,1}^\x{e-imp}} = \frac{2\pi}{\hbar} n_\x{imp} V_\x{cell} \sum_2 \left|g_{12}^\x{imp} \right|^2 \delta^\x{G}_\sigma(\epsilon_1 - \epsilon_2) \frac{\mathbf{v}_1 \cdot \mathbf{v}_2}{v_1v_2},\label{eq:tp_imp}
\end{gather}
where $\mathbf{v}_i$ is band velocity of state $i$ and $v_i$ is its norm.
The thermally-averaged momentum relaxation time $\tau_p$ is then defined as the Fermi-Dirac average of $\tau_{p,1}$.
The scattering times, $\tau_\x{c}$, are calculated the same as \eqs{eq:tp_ph}-\eqref{eq:tp_imp}, but without the final factor $\mathbf{v}_1 \cdot \mathbf{v}_2 / v_1v_2$.

From the dynamics of the density matrix, we then track the spin dynamics via
\begin{equation}
	S_i(t) = \tr{ \hat{s}_i \hat{\rho}(t) },
	\label{eq_spin}
\end{equation}
where $\hat{s}_i$ is the spin operator with $i = x,y,z$.
We determine the spin lifetime by fitting to
\begin{equation}
	S_i(t) - S_i^\x{eq} = \left[ S_i(0) - S_i^\x{eq} \right] \exp(-t/\tau_{\x{s}i}) \cos(\omega t),
	\label{eq_ts}
\end{equation}
where $S_i^\x{eq}$ is the spin polarization at long times, $\omega$ is the spin precession frequency, which may or may not appear depending on the scattering strength, and $\tau_{\x{s}i}$ is the spin lifetime for spins oriented along axis $i$~\cite{Xu2021abinitiodynamics}.

\equ{eq:rho} allows us to accurately describe the impact of phonons and impurities on spin dynamics at the \textit{ab initio} level.
However, we also wish to consider the impact of long-range electrostatic disorder, known as electron-hole puddles.
Arising from charges trapped in the oxide, these puddles have a length scale of tens of nanometers~\cite{Deshpande2009puddles, Xue2011puddles}, and can play an important role in charge scatttering in graphene on SiO$_2$~\cite{Adam2007puddletheory, Klos2011puddletheory}.
Such length scales are beyond the reach of \textit{ab initio} techniques, and thus for this type of disorder we employ a linear-scaling real-space approach capable of handling systems containing many millions of atoms~\cite{Fan2021lsqt}.

In this approach, we calculate the time-dependent and energy-resolved mean square displacement (MSD) and spin polarization of an initial state $\ket{\psi}$,
\begin{gather}
	\x{MSD}(E,t) = \Delta X^2(E,t) + \Delta Y^2(E,t), \\
	\Delta X^2 / \Delta Y^2 (E,t) = \frac{ \braket{\psi_{X/Y}(t) | \delta(E-\hat{H}) | \psi_{X/Y}(t)} }{ \braket{\psi | \delta(E-\hat{H}) | \psi} }, \\
	S_i (E,t) = \frac{ \braket{\psi(t) | \hat{s}_i \delta(E-\hat{H}) | \psi(t)} }{ \braket{\psi | \delta(E-\hat{H}) | \psi} },
\end{gather}
where $\ket{\psi(t)} = \hat{U}(t) \ket{\psi}$, $\ket{\psi_{X/Y}(t)} = [\hat{X}/\hat{Y},\hat{U}(t)] \ket{\psi}$, $\hat{X}$ ($\hat{Y}$) is the position operator along the $x$ ($y$) direction, $\hat{U}(t) = \exp(-\x{i} \hat{H} t / \hbar)$ is the time evolution operator, and $\hat{H} = \hat{H}_0 + \hat{H}_\x{aniso}$ is the TB Hamiltonian of \eqs{eq_ham0}-\eqref{eq_ham1}.

To account for electron-hole puddles, we add an onsite term to the Hamiltonian~\cite{Adam2007puddletheory, Klos2011puddletheory},
\begin{gather}
	\hat{H}_\x{dis} = \sum_{i} \left[ \sum_{j=1}^{\neh} V_j \exp\left(-\frac{|\bm{r}_i-\bm{r}_j|^2}{2\leh^2}\right) \right] \hat{c}_{i}^\dagger \hat{c}_{i},
\end{gather}
which describes a total of $\neh$ Gaussian-shaped puddles, each with random center $\bm{r}_j$, random height $V_j \in [-\veh,\veh]$, and uniform spatial width $\leh$.
To mimic measurements of graphene on SiO$_2$~\cite{Deshpande2009puddles, Xue2011puddles}, we let $\neh / N = 0.0004$, where $N$ is the total number of atoms in the system, $\veh = 50$ meV, and $\leh = 10$ nm~\cite{VanTuan2016puddles}.

The spin lifetime can then be extracted from \eq{eq_ts}, and the momentum relaxation time as
\begin{equation}
	\tau_p(E) = 2 D_\x{sat}(E) / v_\x{F}^2(E),
\end{equation}
where $v_\x{F}$ is the Fermi velocity and $D_\x{sat}$ is the saturated time-dependent diffusivity at long times, with the diffusivity given by $D(E,t) = \frac{1}{4} \frac{\partial}{\partial t} \x{MSD}(E,t)$~\cite{Fan2021lsqt}.

To avoid diagonalization of large matrices, the operators $\hat{U}(t)$ and $\delta(E-\hat{H})$ are expanded as a series of Chebyshev polynomials, and the observables $\Delta X^2$, $\Delta Y^2$, and $S_i$ are reconstructed using the kernel polynomial method~\cite{Fan2021lsqt, Weisse2006kpm}.
For efficient calculation of these observables over the entire Hamiltonian spectrum, we let the initial state $\ket{\psi}$ be a random-phase state that is spin polarized along a preferred direction depending on the component of spin relaxation we wish to observe~\cite{Fan2021lsqt}.

\subsection{Graphene on 2D SiO$_2$}

\begin{figure}[t]
	\includegraphics[width=\columnwidth]{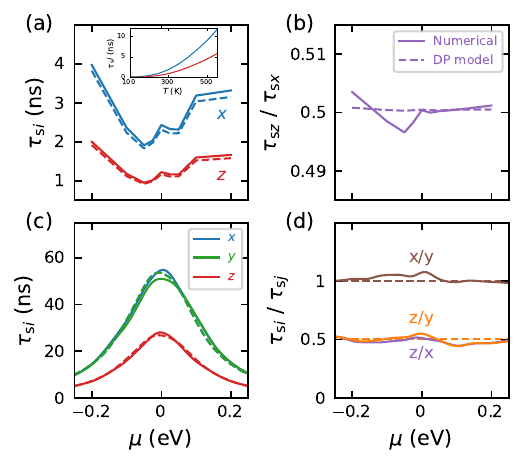}
	\caption{(a) In-plane and out-of-plane spin lifetime as a function of chemical potential for graphene on 2D SiO$_2$, with relaxation driven by electron-phonon scattering at 300 K. Solid lines are from first-principles real-time dynamics, and dotted lines are calculated lifetimes assuming DP spin relaxation from \eq{eq:DP} with \textit{ab initio} parameters. The lifetime as a function of temperature-dependent (at $\mu = 25$ meV above the CB minimum) is shown in the inset. (b) The corresponding spin lifetime anisotropy. (c) Spin lifetime arising from scattering by long-range electron-hole puddles, calculated with the real-space TB approach. (d) The corresponding anisotropies.}
	\label{fig:ts_2d}
\end{figure}

We now examine the nature of spin relaxation in the graphene/SiO$_2$ systems, beginning with the 2D SiO$_2$ substrate.
We consider the impact of electron-phonon scattering using the first-principles density matrix dynamics (FPDMD) approach~\cite{Xu2021}, and electron-hole puddles using the TB model and linear-scaling real-space approach.

The spin lifetime arising from electron-phonon scattering at 300 K is shown in \fig{fig:ts_2d}(a).
The solid lines show numerical results, while the dashed lines show the expected lifetimes assuming a standard D'yakonov-Perel' (DP) mechanism of spin relaxation~\cite{Dyakonov1972, Fabian2007},
\begin{equation}
	\left( \tau_{\x{s}i}^\x{DP} \right)^{-1} = \tau_p \left( \braket{|\bm{\Omega}|^2} - \braket{\Omega_i^2} \right),
	\label{eq:DP}
\end{equation}
where $\bm{\Omega}$ is the $k$-dependent spin-orbit field in units of frequency, and $\braket{...}$ is the average over the Fermi surface. We extract the internal spin-orbit field from the DFT calculations and the momentum relaxation time ($\tau_p$) from Eq.\ \eqref{eq:tp_ph}.
Our calculations indicate that spin relaxation is governed by the DP mechanism arising from substrate-induced inversion symmetry breaking.
Additionally, the spin lifetime increases monotonically with temperature, as shown in the inset of \fig{fig:ts_2d}(a) for a chemical potential of $\mu = 25$ meV.
This increase in $\tau_\x{s}$ corresponds to a monotonic reduction of $\tau_p$ with increasing temperature due to enhanced electron-phonon scattering (see the SM~\cite{SM}), consistent with the picture of motional narrowing in the DP mechanism of spin relaxation.

The spin lifetime anisotropy under electron-phonon scattering, shown in \fig{fig:ts_2d}(b), is $\zeta \approx 1/2$ over the entire range of temperatures and chemical potentials considered.
This behavior is also characteristic of DP spin relaxation resulting from a purely in-plane Rashba‐type spin-orbit field induced by the substrate.

In \figs{fig:ts_2d}(c) and (d), we show the spin lifetimes and anisotropies arising from scattering by electron-hole puddles.
Here we also find that spin relaxation is driven by the DP mechanism.
The opposite dependence on chemical potential arises from the charge scattering; $\tau_p$ exhibits a maximum around the graphene charge neutrality point when driven by electron-phonon scattering, and a minimum when driven by electron-hole puddles.
Overall, our results for graphene on 2D SiO$_2$ echo the behavior that has long been expected for the graphene/SiO$_2$ interface -- spin relaxation driven by the DP mechanism, with a spin lifetime anisotropy of $\zeta = 1/2$ arising from Rashba SOC.
In the next section we see if this behavior extends to the graphene/bulk-SiO$_2$ interface.

\subsection{Graphene on bulk SiO$_2$}

Next we examine spin relaxation in graphene on bulk SiO$_2$.
We consider the structure shown in \fig{fig:structure}(b), a $2 \times 2$ graphene unit cell on a SiO$_2$ slab, which is Si-terminated and passivated with OH (more structural details can be found in the SM Sec.II).
As the phonon modes in this structure are unstable due to the finite thickness of SiO$_2$ slab with a large surface contribution, we instead consider the effect of impurity scattering, as well as electron-hole puddles, on spin relaxation in this system.
For impurity scattering, we consider oxygen vacancies by removing an oxygen atom from the SiO$_2$ layer, relaxing the structure, and then calculating impurity potentials and scattering matrices with defect supercells and employing the FPDMD approach outlined in Sec.~\ref{sec:methodology}.
Meanwhile, electron-hole puddles are considered using the TB model and linear-scaling real-space approach.

\begin{figure}[t]
    \centering
    \includegraphics[width=\columnwidth]{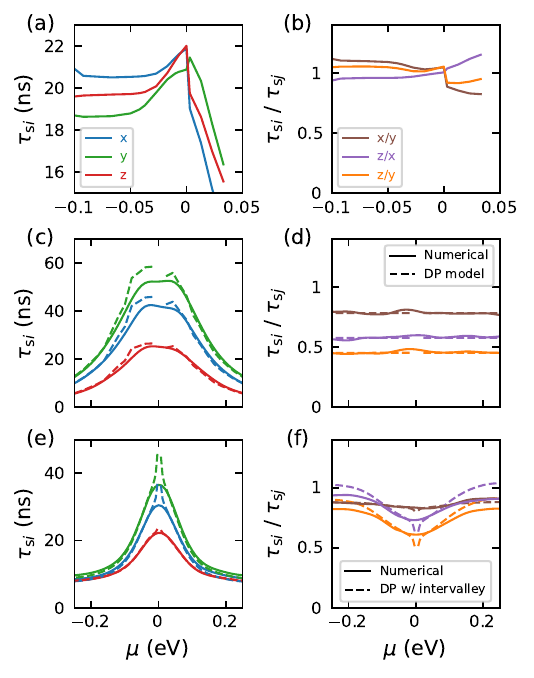}
    \caption{(a) In-plane and out-of-plane spin lifetime as a function of chemical potential for graphene on bulk SiO$_2$, with relaxation driven by oxygen vacancies, and (b) the corresponding spin lifetime anisotropies. (c) Spin lifetime arising from scattering by long-range electron-hole puddles that do not induce intervalley scattering, calculated with the TB model, and (d) the corresponding spin lifetime anisotropies. (e) Spin lifetime arising from short-range electron-hole puddles that induce intervalley scattering, and (f) the corresponding spin lifetime anisotropies.}
    \label{fig:ts_bulk}
\end{figure}

\figu{fig:ts_bulk}(a) shows the spin lifetime as a function of chemical potential, arising from scattering by oxygen vacancies in the SiO$_2$ substrate.
Comparing to \fig{fig:ts_2d}, the spin lifetime is qualitatively different for graphene on 2D SiO$_2$.
Here, $\tau_\x{s}$ is relatively independent of chemical potential, only varying by around $20\%$.
Meanwhile, as shown in the SM~\cite{SM}, the momentum relaxation time varies strongly over this range of $\mu$, by more than a factor of $6$, suggesting that spin relaxation and charge scattering are effectively decoupled.
We examine the mechanism for spin relaxation in further detail in Sec.\ \ref{sec:mechanism} below.

\figu{fig:ts_bulk}(b) shows the spin lifetime anisotropy arising from oxygen vacancies, indicating nearly isotropic ($\zeta \approx 1$) relaxation over the considered range of chemical potentials.
This strongly constrasts with our results for graphene on 2D SiO$_2$, and is more consistent with typical experimental values of anisotropy~\cite{Raes2016, Raes2017, Ringer2018}.

In \fig{fig:ts_bulk}(c) we show the spin lifetime arising from long-range electron-hole puddles.
As before, solid lines are numerical results and dashed lines are the expected lifetimes arising from the DP spin relaxation model of \eq{eq:DP}, indicating a good match.
Panel (d) shows the corresponding spin lifetime anisotropy.
Here we see the appearance of in-plane anisotropy in graphene, with spin lifetimes approximately $20\%$ longer along the $y$-axis than along the $x$-axis.
Meanwhile, $\zeta \approx 1/2$ for both in-plane components, with deviations from $1/2$ arising from the slight in-plane anisotropy.

\figus{fig:ts_bulk}(c) and (d) suggest that graphene on bulk SiO$_2$ behaves like a standard Rashba system in the presence of long-range electron-hole puddles, despite the out-of-plane components in the spin texture shown in \figs{fig:bs_st}(f) and (h).
The reason for this is that, according to our calculations, these long-range puddles do not induce any intervalley scattering.
As first discussed in Ref.~\onlinecite{Cummings2017grTMDC}, a combination of out-of-plane spin polarization and intervalley scattering is required to boost the anisotropy above $1/2$.
To examine this here, we consider a much taller and narrower puddle configuration -- with $\neh/N = 0.0001$, $\veh = 2.7$ eV, and $\leh = 0.435$ nm -- which is representative of charged impurities directly physisorbed on the graphene surface, and is known to induce intervalley scattering~\cite{Cummings2017grTMDC}.

In \figs{fig:ts_bulk}(e) and (f) we show the spin lifetime and anisotropy arising from these short-range impurities.
As shown in panel (f), the out-of-plane to in-plane anisotropy has now risen above $1/2$, ranging between $0.7 < \zeta < 0.95$.
The dashed lines show the spin lifetime and anisotropy expected from a modified DP model of spin relaxation,
\begin{align}
\tau_{\x{s}x}^{-1} &= \tau_p \braket{\Omega_y^2} + \tau_\x{iv} \braket{\Omega_{z}^2} + \tau_p \braket{\Omega_{z\x{,osc}}^2}, \nonumber \\
\tau_{\x{s}y}^{-1} &= \tau_p \braket{\Omega_x^2} + \tau_\x{iv} \braket{\Omega_{z}^2} + \tau_p \braket{\Omega_{z\x{,osc}}^2}, \label{eq:DP_mod}\\
\tau_{\x{s}z}^{-1} &= \tau_p \left( \braket{\Omega_x^2} + \braket{\Omega_y^2} \right ), \nonumber
\end{align}
where $\tau_\x{iv}$ is the intervalley scattering time.
The first two terms in each relaxation rate are the same as originally proposed in Ref.\ \onlinecite{Cummings2017grTMDC} where, owing to time-reversal symmetry, opposite out-of-plane spin polarization in each valley in combination with intervalley scattering leads to a suppression of the in-plane spin lifetime and a consequential increase of the anisotropy.

The third term in the in-plane relaxation rates arises because the out-of-plane spin polarization is nonuniform around the K-point, as shown in \figs{fig:bs_st}(f) and (h).
As the direction of momentum changes with scattering, in-plane spins will experience a fluctuating out-of-plane spin-orbit field even in the absence of intervalley scattering.
This component of the spin-orbit field is given by $\Omega_{z\x{,osc}}(\bm{k}) = \Omega_z(\bm{k}) - \braket{\Omega_z}$.
Because the out-of-plane variation of the spin texture is relatively weak in this system, this extra relaxation term has only a small impact on the spin lifetime.
However, it may play a more important role for other interfaces where this effect is stronger.

\subsection{Spin relaxation mechanism}
\label{sec:mechanism}

Here we delve deeper into the spin relaxation mechanisms at play in the graphene/SiO$_2$ interfaces, and in particular the case with oxygen vacancies in the bulk SiO$_2$ substrate.
Spin relaxation in graphene with SOC is typically attributed to two mechanisms: the Elliott-Yafet (EY) mechanism, where $\tau_\x{s} \propto \tau_p$~\cite{Elliott1954, Yafet1963, Ochoa2012}, and the DP mechanism, where $\tau_\x{s} \propto 1/\tau_p$~\cite{Dyakonov1972, Fabian2007}.
Thus, in a log-log plot of spin relaxation rate vs.\ momentum relaxation rate ($\tau_\x{s}^{-1}$ vs.\ $\tau_p^{-1}$), a slope of $+1$ corresponds to the EY mechanism, a slope of $-1$ corresponds to the DP mechanism, and a slope in between is indicative of a combination of both mechanisms.

In the \textit{ab initio} calculations, we examine this relation by multiplying the scattering matrix $\hat{P}$ of \eq{eq:rho} by a scaling factor $A_\x{scale}$, which directly scales the scattering strength in spin relaxation and the momentum relaxation rate, $\tau_p^{-1} \to A_\x{scale} \times \tau_p^{-1}$.
We can then see how $\tau_\x{s}$ scales with this parameter.
For graphene/2D-SiO$_2$, we consider e-ph scattering at $T=300$ K with the chemical potential $\mu = 25$ meV above the CB minimum.
For graphene/bulk-SiO$_2$, we consider electron-impurity scattering with $\mu = 10$ meV above the CB minimum.
Meanwhile, our TB transport calculations with e-h puddle scattering provide the Fermi energy dependence of both $\tau_\x{s}(E)$ and $\tau_p(E)$, from which we extract the scaling between $\tau_\x{s}^{-1}$ and $\tau_p^{-1}$.

\begin{figure}[t]
    \centering
    \includegraphics[width=\columnwidth]{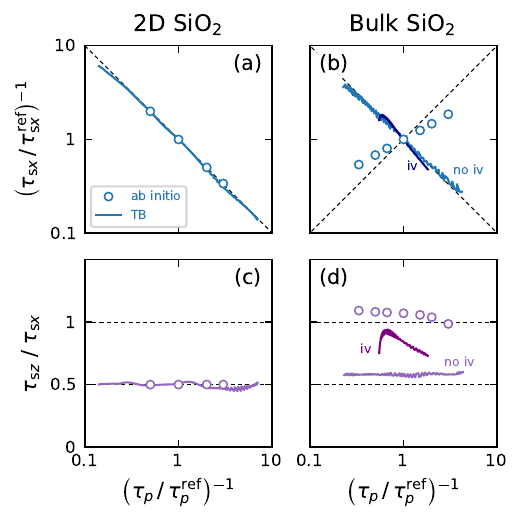}
    \caption{(a,c) Spin relaxation mechanism and anisotropy for graphene/2D-SiO$_2$, and (b,d) the same for graphene/bulk-SiO$_2$. Circles are \textit{ab initio} results, with scattering driven by phonons for Gr/2D-SiO$_2$ and by oxygen vacancies for Gr/bulk-SiO$_2$. Solid lines are from the TB model, with scattering driven by e-h puddles. For e-h puddles in the graphene/bulk-SiO$_2$ interface, we consider puddles with and without intervalley scattering. In panels (a) and (b), the dashed lines with slope $+1$ ($-1$) indicate the scaling arising from pure EY (DP) spin relaxation.
    }
    \label{fig:mechanism}
\end{figure}

This scaling analysis is shown in \fig{fig:mechanism}, where open circles are \textit{ab initio} results and solid lines are from the TB model.
Panel (a) shows the scaling for graphene/2D-SiO$_2$, where the spin and momentum relaxation times have been normalized against the value at $A_\x{scale}=1$ for \textit{ab initio} or a central reference value for TB.
Here, both the \textit{ab initio} and the TB simulations indicate perfect DP scaling, consistent with what we found in \fig{fig:ts_2d} above.
In panel (c), the spin lifetime anisotropy in this system is equal to $1/2$ over the entire range of momentum relaxation, also consistent with the DP mechanism of spin relaxation driven by Rashba SOC.

\figu{fig:mechanism}(b) shows the scaling behavior for graphene on bulk SiO$_2$.
Here, for e-h puddles without intervalley scattering, we see the same behavior as in panel (a), with nearly perfect DP spin relaxation.
For puddles that induce intervalley scattering, the DP relation also holds, with slight deviations arising from a non-constant ratio between $\tau_\x{iv}$ and $\tau_p$; see \eq{eq:DP_mod}.
In panel (d), the anisotropy without intervalley scattering remains slightly above $1/2$ owing to the in-plane anisotropy, while with intervalley scattering $0.7 < \zeta < 0.95$, as already shown in \fig{fig:ts_bulk}.

In contrast, the \textit{ab initio} results show that oxygen vacancies in the bulk SiO$_2$ substrate induces primarily EY-like spin relaxation, with anisotropy $\zeta \approx 1$ for all scattering strengths considered.
This behavior warrants further examination, as it appears to be in contradiction with \fig{fig:ts_bulk}(a) above, where $\tau_\x{s}$ showed little variation with chemical potential $\mu$ despite the scattering time varying dramatically over the same range of $\mu$, suggesting that $\tau_\x{s}$ is decoupled from $\tau_p$.

\begin{figure}
    \centering
    \includegraphics[width=\columnwidth]{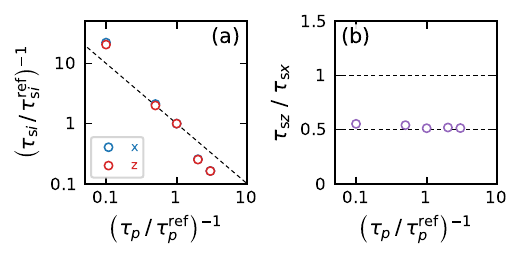}
    \caption{(a) Spin relaxation mechanism and (b) anisotropy for graphene/bulk-SiO$_2$ obtained under the relaxation time approximation of charge scattering.}
    \label{fig:mechanism_rta}
\end{figure}

To examine this further, we replaced oxygen vacancies in bulk SiO$_2$ with generic momentum relaxation using the relaxation time approximation, which amounts to replacing the final term of \eq{eq:rho} with $-(\rho - \rho_\x{eq})/\tau_\x{c}$, where $\tau_\x{c}$ is the carrier relaxation time and $\rho_\x{eq}$ is the equilibrium density matrix.
Here we set $\tau_\x{c}$ to be the same as that arising from oxygen vacancies.
Then we scaled the value of $\tau_\x{c}$ to examine the corresponding variation of the spin lifetime.
The results of this analysis are shown in \fig{fig:mechanism_rta}.
Here we see that in the presence of spin-independent momentum relaxation, spin relaxation in graphene/bulk-SiO$_2$ reverts to the DP mechanism, with an anisotropy slightly above $1/2$, similar to what we see for e-h puddles in the absence of intervalley scattering.

We also consider the spin lifetime and anisotropy expected to arise from the standard EY mechanism.
The spin lifetime is given by $\tau_{\x{s}i}^\x{EY} = \tau_p / \langle b_i^2 \rangle$, where $\langle b_i^2 \rangle$ is the spin mixing of the Bloch states averaged over the Fermi surface~\cite{Elliott1954, Yafet1963, Ochoa2012}, which we extract from the \textit{ab initio} calculations.
We find values of $\tau_{\x{s}i}^\x{EY}$ two to three orders of magnitude larger than those obtained from our first-principles density matrix dynamics simulations, as well as an anisotropy of $\zeta \approx 1/2$, clearly inconsistent with \figs{fig:ts_bulk}(a) and (b).

This indicates that the EY-like spin relaxation seen in \fig{fig:mechanism}(b) is not intrinsic to the spin mixing of the Bloch states of the graphene/bulk-SiO$_2$ system, but rather arises from the specific nature of the scattering event.
We therefore propose that oxygen vacancies induce a local SOC that is responsible for randomizing electron spins during scattering, independent of the momentum relaxation time.
As shown in \eqs{eq:sm_imp} and \eqref{eq:tp_imp}, scaling the strength of the scattering matrix is equivalent to scaling the impurity density, $n_\x{imp} \to A_\x{scale} \times n_\x{imp}$.
This scaling factor then independently scales both $\tau_\x{s}$ and $\tau_p$, yielding the appearance of EY-like spin relaxation.

\section{Summary and conclusion}

In summary, we have presented a comprehensive numerical investigation of spin relaxation and spin lifetime anisotropy in graphene interfaced with 2D and bulk SiO$_2$ substrates.
Through DFT electronic structure calculations, we demonstrated that the graphene/2D-SiO$_2$ interface induces a predominantly Rashba-type in-plane helical spin texture, whereas the graphene/bulk-SiO$_2$ interface exhibits additional in-plane symmetry breaking and nonuniform out-of-plane spin texture.
By constructing an anisotropic TB model, we identified the specific substrate-induced modifications of hopping and spin-orbit coupling that govern these distinct spin textures.

Using first-principles density matrix dynamics simulations, as well as TB transport simulations, we then quantified the effects of electron–phonon scattering, impurity scattering, and electrostatic disorder on the spin relaxation process.
We find that a 2D SiO$_2$ substrate leads to pure DP spin relaxation in the presence of both electron-phonon scattering and electron-hole puddles, with the expected spin lifetime anisotropy of $\zeta = 1/2$.

In contrast, in graphene on a bulk SiO$_2$ substrate we generally find $\zeta > 1/2$.
When electrostatic disorder induces intervalley scattering, the out-of-plane spin texture leads to anisotropies in the range $0.7 < \zeta < 0.95$, consistent with a modified DP mechanism~\cite{Cummings2017grTMDC}.
Meanwhile, oxygen vacancies in the SiO$_2$ substrate lead to primarily EY-like spin relaxation and $\zeta \approx 1$.
This behavior appears to arise from a local variation of the SOC induced in the graphene layer by the defect.

Overall, our results reveal a more complex picture of the graphene/SiO$_2$ interface and its impact on spin relaxation, going beyond the simple Rashba SOC scenario and demonstrating that SOC-driven spin relaxation is capable of yielding spin lifetime anisotropies close to 1.
As today's graphene spintronic devices become cleaner and contact effects are reduced, such behavior may begin to play an important role in the operation of these devices.

\begin{acknowledgments}
This work is primarily supported by the US National Science Foundation under grant number CHE-2505880 and CHE-2203633. Calculations were carried out at the National Energy Research Scientific Computing Center (NERSC), a U.S.\ Department of Energy Office of Science User Facility operated under Contract No.\ DEAC02-05CH11231. This work used the TACC Stampede3 system at the University of Texas at Austin through allocation PHY240212 from the Advanced Cyberinfrastructure Coordination Ecosystem: Services and Support (ACCESS) program~\cite{Boerner2023}, which is supported by US National Science Foundation grants No.\ 2138259, No.\ 2138286, No.\ 2138307, No.\ 2137603, and No.\ 2138296.
ICN2 is funded by the CERCA programme / Generalitat de Catalunya, and is supported by the Severo Ochoa Centres of Excellence programme, Grant CEX2021-001214-S, funded by MCIU/AEI/10.13039.501100011033.
\end{acknowledgments}

\bibliography{ref}

\end{document}